\documentclass[runningheads]{llncs}
\usepackage{graphicx}
\begin{document}
\title{Towards the Artificial Brain:\\A Base Framework for Modelling\\Consciousness and Unconsciousness}
\titlerunning{Towards the Artificial Brain}
%
\author{Daniel Lopes}
\authorrunning{}
%
\institute{
University of Coimbra, \textsc{cisuc}, \textsc{dei}\\
and \textsc{lasi}, University of Minho\\
\email{dfl@dei.uc.pt}
}
\maketitle              
\begin{abstract}
One of the current AI issues depicted in popular culture is the fear of conscious superAIs that try to take control over humanity. And as computational power goes upwards and that turns more and more into a reality, understanding artificial brains is increasingly important to control and drive AI towards the benefit of our societies. This position paper proposes a base framework to aid the development of autonomous, multipurpose artificial brains. To approach that, we propose to first model the functioning of the human brain by reflecting on and taking inspiration from the way the body, consciousness, and unconsciousness interact. To do that, we tried to model events such as sensing, thinking, dreaming, and acting, thoughtfully or unconsciously. We believe valuable insights can already be drawn from the analysis and critique of the presented framework, and that it might be worth implementing it, with or without changes, to create, study, understand, and control artificially conscious systems.


\end{abstract}

\section{Introduction}
Humanity has always been afraid of the unknown. And although that might represent an important survival method, when new technologies are concerned, history seems to continuously suggest that the pros tend to be greater than the cons. For instance, one may take the example of the industrial revolution or the digital revolution. Some types of jobs became obsolete. However, new, more dignifying jobs appeared. Thus, as long as people can learn to adapt, we believe job changes do not need to represent a problem but an opportunity for easier and more dignifying lives in which people can take more time for themselves and create and perform new tasks that they enjoy better or that can elevate their labour to another level.

Nonetheless, it might be consensual that any tool can be used for either beneficial or harmful practices. Thus, putting our efforts into knowing and dominating the diverse existing technologies and, especially, emerging ones is of the utmost importance to put them to work towards the right objectives, as well as knowing how to control them and defend our societies whenever such technologies are used harmfully. 

One of the limelight technologies nowadays is Artificial Intelligence (AI), which is already used in most of the software used for either professional or daily-life purposes. The growing tendency must only escalate with new developments being found every day, coming from every part of the world, making AIs more and more intelligent and in charge of more and more crucial tasks.

In an era where decentralisation is a hot issue, one may wonder about the pros and cons of the most powerful AIs being held by centralised companies, especially when proprietary private code is concerned, i.e., good single leaders can be way more effective than democratic systems; however, bad single leaders can also be way more harmful, especially when controlling powerful tools. 

Therefore, it is of the utmost importance that all the available knowledge on established and emerging technologies is publicly shared. 
Furthermore, we believe AI must still be developed and studied so the maximum number of people can understand and control it, not only to foster market competition and evolve such technologies to solve real problems but also so we can properly legislate about them and join our efforts to develop defense mechanisms against the misuse of such technologies. 

One of the hottest AI issues depicted in popular culture, e.g., by means of movies, has always been the fear of conscious superAIs that try to take control over humanity. And as computational power goes upward, that might become more and more of a reality. Thus, understanding artificial brains is increasingly important to control and drive AI towards the benefit of our societies.
Furthermore, if the scientific community happens to stop developing AI, someone else will do it, and the community will no longer have the knowledge to control it or fight it back, and that might be the real danger to be concerned about. Thus, we believe that still developing, studying, publishing, and legislating AI must always be the way to go.

In that sense, this position paper proposes a base framework to aid the development of autonomous, multipurpose artificial brains. To approach that, we propose to first model the functioning of the human brain by reflecting on and taking inspiration from the way the body, consciousness, and unconsciousness interact. To do that, through introspective observation, we tried to model events such as sensing, thinking, dreaming, and acting, thoughtfully or unconsciously. 

Although this is a preliminary proposal, we believe valuable insights can already be drawn from the analysis and critique of the presented framework and that it might be worth taking it as a base, with or without changes, to create, study, understand, and control artificially conscious systems.



\section{Proposed Framework}

\begin{figure}
    \centering
    \includegraphics[height=\textheight]{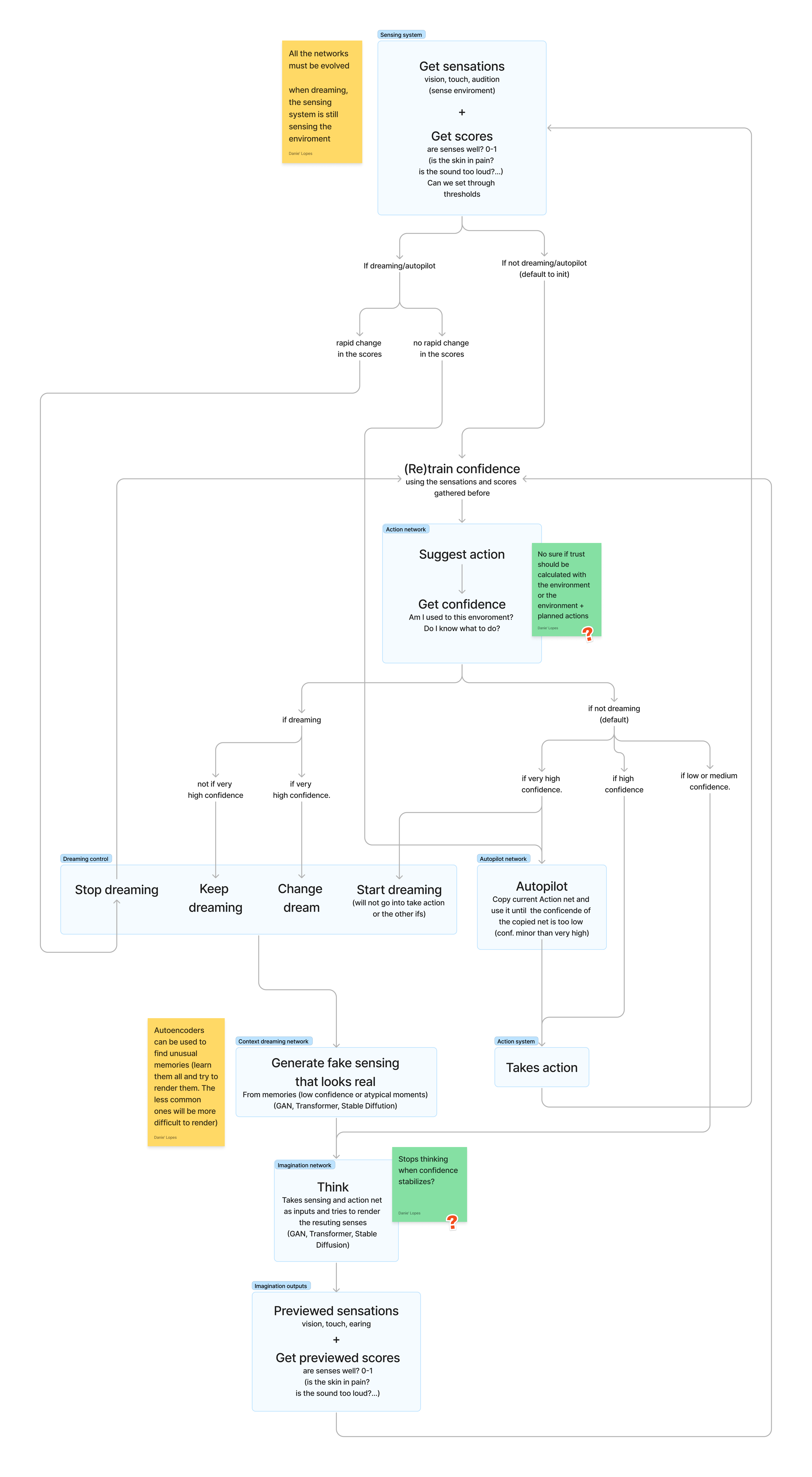}
    \caption{Schematic representation of the proposed framework.}
    \label{framework}
\end{figure}

Figure \ref{framework} illustrates the proposed framework. See the last updated scheme in detail on the following link
\textit{https://www.figma.com/file/3eyyDij8mPgE7eLy6lGjem/ Towards-the-Artificial-Brain?node-id=0\%3A1\&t=dm8FKe0whabctKiz-1}.

We highlight that the presented framework was first thought to be used to build artificial robot brains. Also, we propose to evolve all the networks referred to, i.e., train them using evolutionary computation \cite{bentley}.

The framework starts with a module named \textit{sensing system} that is responsible for collecting inputs from the surrounding environment, as existing robots frequently do. For example, this can include vision, touch, or audio inputs. We refer to these values as direct inputs. Also, such direct inputs must be evaluated to create additional values relative to whether the sensors are being ``injured'', e.g., whether the sound is too loud or something is damaging the robot's ``skin'' or any other body features. The same is valid for whether positive values are being received, e.g., ``pleasure'' values. Such ``pain''/``pleasure'' values can be set, for example, through thresholds. Direct inputs and ``pain''/``pleasure'' values together make the outputs of the sensing system.

Following the sensing system, there are two possible states to continue with, which refer to whether the system is dreaming or not. The latter, not dreaming, is the default state in which one must initiate the system. The not dreaming state means nothing needs to be done so far, so one can advance in the pipeline by training the network we refer to as \textit{action network} using the outputs of the \textit{sensing system}. This network must be able to propose the next actions of the robot by suggesting an input to each of the robot's actuators, e.g. in what position its arms, legs, or head must be, or even what sound to make. Additionally, it must preview a confidence value for whether the action will be successful, i.e., is the system used to the current situation? Does it know what to do? Will this action result in good or bad ``pain''/``pleasure'' values?

Again, following the \textit{action network} module, the dreaming or not dreaming states are the two possible ways to continue. And again, not dreaming is the default.
By continuing with the not dreaming state, one out of three scenarios is possible depending on whether the confidence value returned by the \textit{action network} was (i) low/medium, (ii) high or (iii) very high (can be defined using a threshold).

In case (ii), high confidence, i.e., the system is secure about the success of its next action, the outputs suggested by the \textit{action network} will be passed to the \textit{action system} which, in turn, will pass the suggested values to the robot's actuators, making the action happen in the real environment, e.g., moving an arm or making some sound.

In case (i), low/medium confidence, the outputs from the \textit{sensing system} as well as the \textit{action network} must be fed together as inputs to the \textit{imagination network} which is responsible for ``thinking''. More specifically, the inputs must be used to try to render the resulting senses as realistically as possible, i.e., what the robot should be seeing, hearing or touching (or other senses) after the suggested actions. To do that, for example, Generative Adversarial Networks (GANs) \cite{gans} or Transformers \cite{transformers} can be used, or new techniques can be developed or evolved. Thereafter, ``pain''/``pleasure'' values must be assessed using the same method/thresholds as in the \textit{sensing system}. The rendered senses and the ``pain''/``pleasure'' values must then be fed as inputs to the \textit{action network} to retrain it, and the pipeline goes again from there on.

Lastly, in case (iii), for very high confidence, the system continues to the \textit{autopilot network} which must copy the current \textit{action network} and use it to take actions (as the \textit{action system} does) until the confidence value retrieved by the copied \textit{action network} gets lower than very high. We refer to this process as \textit{autopilot}. Unless there is a rapid change in the ``pain''/``pleasure'' values or unless these are warning, after taking action, the system will proceed to the \textit{sensing system} again, and the outputs of the latter will be fed directly to the \textit{autopilot network}, so the system can still be taking actions automatically.

In parallel to the \textit{autopilot network} taking actions, the system must start a new thread we refer to as \textit{dreaming}. If there is a rapid change in the ``pain'' / ``pleasure'' values or if the values are warning, the system will end \textit{dreaming} and the pipeline goes from there on as previously explained. 

The \textit{dreaming} thread is meant to fake sensing inputs in a way that makes them look real. For example, out of memories, such as saved senses referring to low confidence or atypical moments. In a more practical way, this task can comprise generating an image to replace vision, respective sounds to replace audition, and respective touching senses, i.e., to sense a fake new environment based on previous memories, just like in a dream. Again, this task could be accomplished using GANs, Transformers or Stable Diffusion techniques. The generated fake senses must then be fed to the \textit{imagination network} and the pipeline must proceed from there on, as previously explained. However, in this next loop, after the \textit{action network}, the system will not continue to the \textit{not dreaming} condition but the \textit{dreaming} one instead.

By proceeding to the \textit{dreaming} condition, two other states are possible: the \textit{action network} is either ``very high'' confident of the suggested actions or not. If not, the system continues dreaming by evolving the same scenario (e.g., creating congruent next frames). If the system is already ``very high'' confident of the actions to take (e.g., during a certain period of time), the dream must be changed by generating new fake sensing inputs, unrelated to previous ones, and the pipeline must proceed as previously explained.

We believe the implementation of this pipeline (or ones based on the main ideas behaving it) can lead to autonomous learning, e.g., aiding the system to retrieve learning feedback from the environment through its senses, as biological beings do. Also, we believe the capacity to dream of unseen environments can aid the development of the system's abilities and lead the system to be artificially creative.

However, notice that this is a work in progress and, currently, it is not based on scientific experiments but rather on thoughtful analysis. Implementation and testing are needed to prove its value.

\bibliographystyle{splncs04}
\bibliography{bib}

\end{document}